\documentclass[fleqn,usenatbib]{mnras}
\usepackage{newtxtext,newtxmath}
\usepackage{natbib}
\usepackage{graphicx}
\usepackage{color}
\usepackage{epsfig}
\usepackage{mathrsfs}
\usepackage{ulem}
\usepackage{times} 
\usepackage{balance} 
\usepackage{longtable,lscape}
\usepackage{url}
\usepackage{bm}

\usepackage[T1]{fontenc}

\usepackage[switch]{lineno}
\usepackage{color}
\usepackage{mathrsfs}

\newcommand{\bhm}{M_{\bullet}}
\newcommand{\calD}{{\cal D}}
\newcommand{\civ}{{\sc C iv}}
\newcommand{\dotm}{\dot{m}}
\newcommand{\ergs}{\rm erg\,s^{-1}}

\newcommand{\feii}{Fe {\sc ii}}

\newcommand{\mgii}{Mg {\sc ii}}
\newcommand{\mathdotM}{\dot{\mathscr{M}}}
\newcommand{\Rtr}{R_{\rm tr}}
\newcommand{\oiii}{[O {\sc iii}]}
\newcommand{\sunm}{M_{\odot}}

\newcommand{\Rg}{R_{\rm g}}

\begin{document}

\def\ihep{Key Laboratory for Particle Astrophysics, Institute of High Energy
Physics, Chinese Academy of Sciences, 19B Yuquan Road, Beijing 100049, China}

\def\ucas{School of Astronomy and Space Science, University of Chinese Academy of
Sciences, 19A Yuquan Road, Beijing 100049, China}

\def\naoc{National Astronomical Observatories of China, The Chinese Academy of
Sciences, 20A Datun Road, Beijing 100020, China}

\title[{Final stage of merging binaries of supermassive black holes:}]
{Final stage of merging binaries of supermassive black holes: observational signatures}

\author[J.-M. Wang et al.]
{Jian-Min Wang$^{1,2,3}$\thanks{E-mail:wangjm@ihep.ac.cn},
Yu-Yang Songsheng$^{1,2}$, Yan-Rong Li$^{1}$ and Pu Du$^{1}$\\
$^{1}${\ihep}\\
$^{2}${\ucas}\\
$^{3}${\naoc}}

\date{}
 
\pagerange{\pageref{firstpage}--\pageref{lastpage}} \pubyear{2020} 
\maketitle
\label{firstpage}

\begin{abstract}
There are increasing interests in binary supermassive black holes (SMBHs), but 
merging binaries with separations smaller than $\sim 1\,$light days ($\sim 10^{2}\,$
gravitational radii for $10^{8}\sunm$), which are rapidly evolving 
under control of gravitational waves, are elusive in observations. In this
paper, we discuss fates of 
mini-disks around component SMBHs for three regimes: 1) low rates 
(advection-dominated accretion flows: ADAFs); 2) intermediate 
rates; 3) super-Eddington accretion rates. Mini-disks with intermediate rates are 
undergoing evaporation through thermal conduction of hot corona forming a hybrid radial 
structure. When the binary orbital periods are shorter than sound propagation timescales 
of the evaporated mini-disks, a new instability, denoted as sound instability, arises 
because the disks will be highly twisted so that they are destroyed. We demonstrate a 
critical separation of $A_{\rm crit}(\sim 10^{2}\Rg)$ from the sound instability of the 
mini-disks and the cavity is full of hot gas. For those binaries, component 
SMBHs are accreting with Bondi mode in the ADAF regime, showing periodic variations 
resulting from Doppler boosting effects in radio from the ADAFs due to orbital motion. 
In the mean while, the circumbinary disks (CBDs) are still not hot enough (ultraviolet 
deficit) to generate photons to ionize gas for broad emission lines. For slightly 
super-Eddington accretion of the CBDs, \mgii\, line appears with decreases of UV deficit, 
and for intermediate super-Eddington Balmer lines appear, but \civ\, line never unless 
CBD accretion rates are extremely high. Moreover, if the CBDs are misaligned 
with the binary plane, it is then expected to have optical periodical variations with 
about ten times radio periods. 
\end{abstract}

\begin{keywords}
{galaxies: active -- quasars: general – quasars: supermassive black holes}
\end{keywords}

\section{Introduction}
Searching for binaries of supermassive black holes (SMBHs) in galactic centers draw much 
attention not only for galaxy formation and evolution but also for detection of low-frequency 
gravitational waves through Pulsar Timing Array (PTA). As natural consequence of galaxy 
mergers, binary SMBHs evolve through different phases 
\citep[e.g.,][]{Begelman1980,Milos2001,Volonteri2009}, however, those close binaries of 
SMBHs (CB-SMBHs: separations less than 0.1\,pc) are still elusive in observations 
\citep{Wang2020a,Bogdanovic2021}. 
There are many characteristics of CB-SMBHs from broad emission 
lines \citep{Popovic2012,DeRosa2019,Bogdanovic2021,Charisi2022} even more clues from
observational evidence from shifting broad emission line, mass deficits in density 
profile of galactic nuclei, UV continuum deficit, reverberation mapping signals to interferometric 
signals of GRAVITY/VLTI, gas dynamics of circumbinary disks, to polarized spectra \citep{Wang2020a},
but it is hard to draw conclusive remarks on individual candidates \citep{Wang2020a}.
Difficulties of identifying CB-SMBHs rest on lack of conclusive criterions.  
What are the key features of binary SMBHs at different stages in light of separations?

When the binary spirals in a stage with separations of $\sim 10^{4}\Rg$ 
(where $\Rg=1.5\times 10^{13}\,M_{8}\,{\rm cm}$ is gravitational radius, $M_{8}$ is 
SMBH mass in units of $10^{8}\sunm$), the configurations of the
binary SMBHs with accretion disks and broad-line regions (BLRs) are still independent 
of each other according to the $R-L$ relation \citep{Kaspi2000,Bentz2013,Du2019}.
In this scale of separations, two SMBHs have their own BLRs and mini-disks \citep{Shen2010}. 
Identifying these binary SMBHs can be done by examining 2D transfer functions from 
reverberation mapping campaigns \citep{Wang2018,Songsheng2020,Kovacevic2020a,Ji2021}. 
The 2D transfer functions of binary BLRs rotating around each other will deliver 
binary information of the orbital 
motion \citep{Wang2018}, which are obviously different from a single 
BLR \citep[e.g.,][]{Welsh1991}. A dedicated RM campaign of Monitoring AGNs with H$\beta$ 
Asymmetry (MAHA) for this goal has been undergoing through Wyoming Infrared Observatory 
(WIRO) 2.3m telescope \citep{Du2018,Brotherton2020}. Being different from this approach, 
long term campaigns of monitoring variations of profiles proceeds smoothly to test 
periodical shifts of red and blue peaks 
\citep[]{Montuori2012,Decarli2013,Runnoe2017,Doan2020,Guo2019,Popovic2021}.
Moreover, GRAVITY/VLTI (Very Large Telescope Interferometer) with unprecedentedly 
high spatial resolution ($\sim10\mu$as) has been applied to measure kinematics and spatial 
distributions of ionized gas of AGNs \citep[]{GC2017,GC2018,GC2020,GC2021}. This
powerful tool has been first suggested by \cite{Songsheng2019} and \cite{Kovacevic2020b} to 
apply to close binaries of SMBHs. Combining SpectroAstrometry (SA) and RM (hereafter SARM)
offers an opportunity to identify them and measure orbital parameters of the binaries for 
testing nano-Hz gravitational waves, which is similar to that for cosmic distances
\citep{Wang2020c}.

\begin{figure}\label{fig:cartoon}
\centering
\includegraphics[angle=0,width=0.4\textwidth,trim = 250 80 285 40, clip]{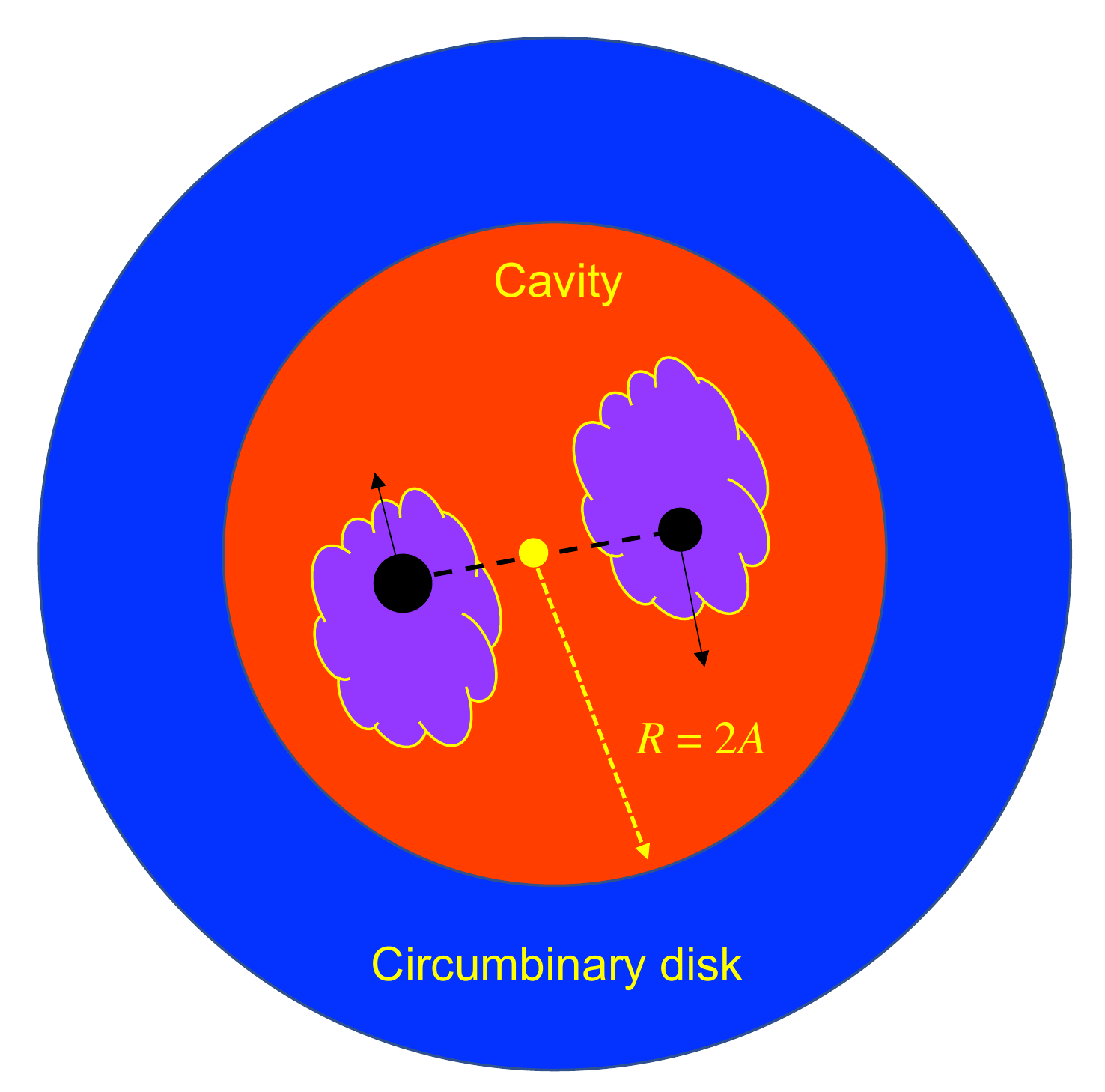}
\caption{Carton of a merging binary of SMBHs with a circle orbit 
(separation is about $10^{2}$ gravitational radius). Mini-disks around SMBHs are 
not able to survive so that a cavity is full of hot gas due to interaction with 
the binary. A circumbinary disk is then formed with a radius $R\approx 2A$. Each 
SMBH is accreting through Bondi mode from the cavity. Since SMBH Bondi accretion 
has low accretion rates, periodical variabilities in radio are expected due to 
orbital motion of the binary. See details in main text.}
\end{figure}

Periodicity of AGNs receives much attention for candidates of CB-SMBHs in recent years.
If the periodicity is caused by the binary SMBHs, separations of binaries with periods
of a few years are about $\sim 10^{3}\,\Rg$.
There are several long term campaigns of photometric surveys in time domain such as 
the Catalina Real-time Transient Survey (CRTS), Palomar/Zwicky Transient Factory (P/ZTF),
Panoramic Survey Telescope and Rapid Response System (Pan-STARRS) and Dark Energy Survey 
(DES) for specific types of astronomical transients.
There is growing evidence for periodical variations of optical light curves of AGNs from 
$\sim 10$ year photometric campaigns for CB-SMBHs 
from CRTS \citep{Graham2015b}, from PFT \citep{Charisi2016}, from PanSTARRS \citep{Liu2019},
from DES \cite{Chen2020}. Some of these periodic AGNs are likely not binary SMBHs since 
red-noise variabilities can generate periodic-like variations 
\citep[e.g.,][]{Vaughan2016}. There are also a few individual periodic AGNs, such as
OJ 287 \citep{Valtonen2008}, PG\,1302-102 \citep{Graham2015a}, NGC 5548 \citep{Li2016} 
and Akn 120 \citep{Li2019}. Extremely eccentric CB-SMBHs with $10^{3}\Rg$ separations 
merging fast may lead to changing-look events of AGNs \citep{Wang2020b}, however, such
separations are still too large for GW to test gravitational waves in the chirping phase.
     
In this paper, we investigate the final stage of close binaries of SMBHs whose orbits
are completely controlled by gravitational waves ($\sim10^{2}\,\Rg$). 
We show major physics of accretion onto 
such a close binary of SMBHs, and the peculiar spectral energy distributions of continuum.
Moreover, the Doppler boosting effects due to fast orbital motion can be observed, and 
polarization observations are more useful \citep{Dotti2022}.
Spectral features of CB-SMBHs at the final stages are determined by the accretion physics.
We predict their observational appearance.

\section{Merging Binaries: a critical separation}
\subsection{Fiducial numbers}
In this paper, for a simplicity, we employ the third Kepler's law to approximate the 
orbits of merging binary SMBHs. The period is given by
\begin{equation}\label{Eq:Porb}
P_{\rm orb}=0.1\,a_{2}^{3/2}M_{8}(1+q)^{-1/2}\,{\rm yr},
\end{equation}
where $M_{8}$ is the primary SMBH mass, $a_2=A/10^2\,\Rg$ is the separation 
of the binary SMBHs, $\Rg=G\bhm/c^{2}$ is the gravitational radius, $G$ is the gravitational 
constant, $c$ is speed of light,  and $q$ is the mass ratio of the secondary to the primary. 
The orbital decay timescale due to gravitational waves is given by 
\begin{equation}\label{eq:taugw}
\tau_{\rm gw}=\frac{5}{256}\frac{c^{5}A^{4}}{G^{3}\mu\bhm^{3}}
             =30.5\,a_{2}^{4}\mu^{-1}M_{8}\,{\rm yr},
\end{equation}
for circle orbits from \cite{Peters1964}, where $\mu=q(1+q)$. The GW radiation timescale 
is very sensitive to the separation. The intrinsic strain amplitude is
\begin{equation}
h_{s}=3.2\times 10^{-17}\,q(1+q)^{-1/3}P_{0.1}^{-1}d_{100}^{-1}M_{8}^{5/3},
\end{equation}
with frequencies of $\sim 10^{-6}\,$Hz, where $P_{0.1}=P_{\rm obs}/0.1\,{\rm yr}$, 
$d_{100}=d_{\rm L}/100\,{\rm Mpc}$ is the distances to observers in units of 100\,Mpc. 
The GW frequency and the strain are detectable by future PTA of Square Kilometer Array 
(SKA) and PTA facilities. The period decays due to gravitational waves from a circle 
orbital binary is given by
\begin{equation}\label{eq:rates}
\left(\frac{dP_{\rm orb}}{dt}\right)_{\rm gw}=-0.3\,q(1+q)^{1/2}a_{2}^{-5/2}\,{\rm day\,yr^{-1}},
\end{equation}
where we make use of $(dP_{\rm orb}/dt)_{\rm gw}\approx \left(3P/2A\right)\left(dA/dt\right)$. 
We note that this rate in this approximation is independent of SMBH masses.
Sophisticated treatment involves post-Newtonian approximations \citep[e.g.,][]{Blanchet2014}, 
which is beyond the scope of this Letter. Such a decaying rate can be easily detected by 
monitoring campaigns.

As we show below, fates of the mini-disks around component SMBHs are determined by the
orbital motion of the binaries when their separations are close enough. During the final stage, 
accretion onto the binaries is competing with orbital evolution governed by gravitational 
waves. If we can identify one target at the final stage,  the periodic variabilities 
can be found, and period changes can be measured accurately enough for observational appearance 
owing to the sound instability.

\subsection{A critical separation of binaries}
It has been well understood that a cavity with a radius of $R\approx 2A$ will be formed 
at the center of the circumbinary disk (CBD) after the pioneering
work of \cite{Artymowicz1994}. Many numerical simulations have been done for the 
configurations of accretion onto binary SMBHs. Some results from numerous simulations 
are well established, in particular, 3D-GRMHD simulations showed the mini-disks around 
the component SMBHs exist inside the cavity, and they could have different accretion 
rates \citep{Farris2011,Noble2012,Giacomazzo2012,Gold2014,Zilhao2015,Bowen2018,Bowen2019,
Armengol2021,Cattorini2021,Combi2021,Noble2021,Paschalidis2021,Gutierrez2022}. Ultraviolet
deficit appears as the major features of spectral energy distributions of the binaries
\citep[e.g.,][]{Gultekin2012,Sesana2012,Gutierrez2022}, however, this well-known feature
remains highly uncertain in observations because of its uniquity (dusty reddening and 
extinction can easily lead to a deficit similar to this effect, see \citealt{Leighly2016}).

In order to conveniently discuss fates of the mini-disks, we define   
dimensionless accretion rates of the CBD as 
$\mathdotM=\dot{M}_{\rm CBD}/\dot{M}_{\rm Edd}$ from the mass rates of $\dot{M}_{\rm CDB}$,
where $\dot{M}_{\rm Edd}=L_{\rm Edd}/c^{2}$ and $L_{\rm Edd}$ 
is the Eddington luminosity for each component of binary SMBHs. 
Component SMBHs are accreting gas supplied by the CBD, and 
the primary has a rate $\dot{M}_{\bullet}=f_{\rm p}\dot{M}_{\rm CDB}$, where $f_{\rm p}$ 
is a fraction of the CBD rates. We use
$\dot{m}=\dot{M}_{\bullet}/\dot{M}_{\rm Edd}$ as the dimensional
rates of the mini-disk around the primary SMBH. The properties of the mini-disks as
a key to examine the observational features are insufficiently understood
though there are some simplified discussions in the scheme of standard accretion disk model 
\citep[e.g.,][]{Haiman2009}.

\subsubsection{Mini-disks of SMBHs}
Outer radii of mini-disks are constrained by their Roche lobes,
$R_{\rm out}/A\approx 0.49q^{2/3}/\left[0.6q^{2/3}+\ln(1+q^{1/3})\right]$ 
\cite[e.g.,][]{Eggleton1983}. In light of $\mathdotM$, we discuss the following cases for
mini-disks of accretion onto component SMBHs. Within $R_{\rm out}$, the disks are mainly 
governed by the single component SMBH whereas gas beyond $R_{\rm out}$ spread into the 
cavity. Gas supply to the component SMBHs is mainly determined by the interaction between 
the binary and the CDB, the factor $f_{\rm p}$ is beyond discussions in this paper.

{\textit{Case A}:} CBDs have rates of 
$\mathdotM\lesssim\mathdotM_{c}\approx 0.25\,\alpha_{0.3}^{2}$ 
\citep[see Eqn.\,52 in][]{Mahadevan1997}, where $\alpha_{0.3}=\alpha/0.3$ is the viscosity
parameter in advection-dominated accretion flows (ADAFs) \citep[][]{Narayan1994}.
The entire system is in ADAF regime, which is characterized by hot temperature (close 
to virial 
ones) and inefficient radiation. The outer boundary of the ADAFs is determined by the
evaporation-driven truncated radius (see Case B for a brief introduction of evaporation).
\cite{Meyer2002} and \cite{Czerny2004} approximated the truncated radius of 
$r_{\rm ADAF}^{\rm out}=r_{\rm evap}\approx378.6\,\mathdotM_{0.1}^{-0.85}\beta_{0.5}^{2.5}$
in units of $\Rg$ whereas $r\ge r_{\rm ADAF}^{\rm out}$ disks follow solution of middle
regions of the \cite{Shakura1973}. As we shown in Case B, 
$r_{\rm ADAF}^{\rm out}\gtrsim a_{\rm crit}$
implies that evaporation dominates over the sound instability and the cavity is naturally
full of hot gas.

Sound speed of ADAFs is faster than (or comparable to) the orbital motion of binary 
SMBHs, and hence ADAFs follow the SMBHs. Spectral energy distributions of ADAFs have 
been well understood. They are composed of several bumps from radio to hard X-rays 
due to synchrotron and multiple inverse Compton scatterings, respectively 
\citep[e.g.,][]{Manmoto2000}. Figure \ref{fig:sed} shows SEDs of ADAFs around SMBHs. 

AGNs with low accretion rates usually appear as low-ionization nuclear emission-line 
regions (LINERs) \citep[e.g.,][]{Ho2008}. If one LINER contains merging binary of SMBHs,
as we show in \S\ref{sec:Doppler}, radio and X-ray emissions from the binary ADAFs show 
periodic variations due to Doppler boosting. Periodicity of radio and X-ray light curves
will be reliable diagnostic of merging binary SMBHs. On the other hand, the transition 
regions from Shakura-Sunyaev disk to ADAF could radiate some UV photons 
\citep[see SEDs in][]{Liu2022}, which are photonionizing 
the CBDs, emerging as weak broad-line emitters. 
 
Moreover, binary SMBHs, if existing in weak galactic nuclei, will show periodic variations
in radio bands from binary ADAFs. SMBHs in normal galaxies are known to undergo Bondi 
accretion from the interstellar medium (ISM) evidenced by radio emissions 
\citep{Franceschini1998,Nyland2016,Grossova2022} and X-rays of {\textit{Chandra}} 
observations 
\citep[e.g., $\mathdotM\sim 10^{-2}$ in NGC\,6166 from][]{DiMatteo2001,Grossova2022}. 
Radio and X-ray monitoring 
campaigns for periodic variations will provide a robust way of identifying binary SMBHs 
in normal galaxies. Candidates can be selected from galaxies either with merger signatures 
(tails), or with 
mass deficit in nuclear density profiles \citep[e.g.,][]{Ebisuzaki1991}. These targets
are suggested to the project of the Karl G. Jansky Very Large Array Sky Survey (VLASS) 
\citep{Lacy2020}, which is designed for time domain survey of radio sources. 
X-ray monitoring campaigns will also be useful to explore periodicity due to binary 
SMBHs since X-rays are less contaminated by stars in host 
\citep[see the review of hot gas in normal galaxies by][]{Mathews2003}.

{\textit{Case B}:} CBDs have intermediate $\mathdotM$ in the regime of the standard 
accretion disk \citep{Shakura1973}. Fates of the mini-disks depend on their radial 
structures. The classical model in this regime has three distinct regions: 1) inner 
region with a radius of $r\le r_{\rm inn}= 167.1\,(\alpha_{0.1}M_{8})^{2/21}\dotm^{16/21}$ 
dominated by radiation pressure and electron scattering, where $\alpha_{0.1}=\alpha/0.1$; 
2) middle region with a radius 
of $r_{\rm inn}\le r\le r_{\rm mid}= 2.5\times 10^{3}\dotm^{2/3}$ dominated by gas 
pressure and electron scattering and 3) outer regions ($r\ge r_{\rm mid}$) dominated 
by gas pressure and absorption (free-free, free-bound and bound-bound) 
\citep[see extensive discussions in][]{Kato2008}. In this regime, 
the viscosity parameter $\alpha$ is significantly smaller 
than that in ADAFs \citep[e.g.,][]{Mahadevan1997}. The inner part releases most of 
gravitational energy, however, it turns out that the model of this part should be 
revised by considering more sophisticated physics. 

Hot corona above the cold disk generally exists evidenced by X-ray emissions in AGNs 
\citep[e.g.,][]{Haardt1991}. It plays a key role in the radial structure of the cold 
disk through very efficient evaporation \citep[e.g.,][]{Meyer1994,Liu1999}.
In such a scenario, the inner part becomes ADAF, where gas pressure dominates over the
radiation. Evaporation
of the cold part depends on viscosity, magnetic fields and Coulomb coupling between 
electrons and protons in the hot corona \citep[see a recent review of][]{Liu2022}.
Considering the roles of magnetic fields on the suppression of evaporation, \cite{Meyer2002}
made numerical calculations and \cite{Czerny2004} obtained an approximate expression of
evaporation rates in units of $\dot{M}_{\rm Edd}$
\begin{equation}
\dotm_{\rm evap}=7.1\,\beta_{0.5}^{2.94}r_{1}^{-1.17},
\end{equation} 
converted from the truncated radius ($\Rtr$) given by Eqn.(10) in \cite{Czerny2004}, 
where $\beta_{0.5}=\beta/0.5$ and
$\beta$ is the ratio of gas pressure to the total of gas and magnetic fields. 
Evaporation depends on magnetic fields because the thermal conduction coefficient
can be significantly suppressed by the fields. $\beta=0.5$ implies a magnetic field
in equipartition with the thermal gas in hot corona.
The evaporation timescale is given by
\begin{equation}
\tau_{\rm evap}=\frac{\pi R^{2}\Sigma}{\dotm_{\rm evap}}\approx 
0.23\,(\alpha_{0.1}\dotm)^{-1}\beta_{0.5}^{-2.94}M_{8}r_{1}^{4.67}f^{-1}\,{\rm yr},
\end{equation}
where $\Sigma$ is the surface density taken from Eqn.(\ref{Eq:surface}), 
$f=1-(r_{\rm in}/r)^{1/2}$ is
the inner boundary factor and $r_{\rm in}$ is the inner radius of accretion disks. 
It should be noted that $\tau_{\rm evap}$ is very sensitive to radius.
This timescale is comparable with orbital periods of merging binary SMBHs 
(see Eqn.\,\ref{Eq:Porb}).

%
%

In the inner region, the sound propagation of the mini-disks
along $r$-direction is given by $\tau_{\rm sound}=R/c_{s}$, where 
$c_{s}\approx(P_{\rm rad}/\rho)^{1/2}$ is sound speed, the radiation pressure of
$P_{\rm rad}=aT_{c}^{4}/3$ dominates over gas pressure, where $a$ is the black body 
constant and $T_{c}$ is the temperature of the mid-plane (given in Appendix).
For a simplicity, we still use the solution of \cite{Shakura1973} model for regions 
beyond $\Rtr$, we have
%
%
%
\begin{equation}
\tau_{\rm sound}=
 0.058\,\dotm^{-1} M_{8}r_{1}^{5/2}f_{0.23}^{-1}\,{\rm yr},
\end{equation}
where $c_{s}=8.3\times 10^{7}\dot{m} r_{1}^{-3/2}f_{0.23}\,{\rm cm\,s^{-1}}$ in the 
inner region, $r_{\rm in}=6$ is the last stable radius of the mini-disks for 
Schwartzschild black holes, $f_{0.23}=f/0.23$ at $r=10$. The inner part of the standard 
accretion disk model is divided into two parts: 1) an ADAF in the innermost region within 
$\Rtr$ and 2) a truncated cold part beyond $\Rtr$. We note that the viscosity timescale
$\tau_{\rm vis}\approx R/\upsilon_{r}
=15.5\,\alpha_{0.1}^{-1}\mathdotM^{-2}r_{1}^{7/2}f_{0.23}^{-2}\,$yr, which is much longer
than $\tau_{\rm sound}$ and $P_{\rm orb}$, but comparable with $\tau_{\rm gw}$. 

{\textit{Case C}:} CBDs have super-Eddington accretion rates with $\mathdotM\gg 1$ and the 
mini-disks hence have $\dot{m}\gg1$, which is characterized by photon trapping and transonic 
radial motion \citep{Abramowicz1988,Czerny2019}. Indeed, they exist in the Universe from reverberation 
mapping campaigns of super-Eddington accreting massive black holes (SEAMBHs) selected from
AGNs with strong optical \feii\ and \oiii\ lines
\citep{Du2014,Du2015,Du2018,Du2019}. The highest $\mathdotM$ detected so far
is close to $\sim 10^{3}$ from the SEAMBHs project \citep{Du2018,Du2019}. If 
the CBDs have super-Eddington rates, mini-disks are expected to share a super-Eddington 
rate ($f_{\rm p}\lesssim 1$). As accretion rates increase, the 
mini-disks are insufficiently evaporated by the hot corona, and the evaporation radius is
about $\Rtr\sim 6\,\alpha_{0.1}^{4}(\dot{m}/10^{2})^{2}$ \citep[see Eqn.8 in][]{Czerny2004} 
approaching to the last stable orbit. ADAF disappears in super-Eddington cases evidenced 
by weak corona in high Eddington AGNs \citep[e.g.,][]{Wang2004}. We use the self-similar 
solution of super-Eddington accretion flows from \cite{Wang1999}, and have
\begin{equation}
\tau_{\rm sound}\approx1.1\times 10^{-3}\,M_{8}r_{1}^{3/2}\,{\rm yr}. 
\end{equation}
In the super-Eddington cases, the sound timescale is much shorter than that in sub-Eddington 
accretions.

In the context of binary SMBHs, there are three competing processes governing
mini-disks characterized by: 1) evaporation controlled by hot corona; 2) rapid rotation 
(and evolution) of orbital motion and 3) sound propagation determining
the radial recovery after a perturbation. The fates of the mini-disks are controlled
by the competitions among the timescales, in particular, the presence of binary SMBHs
will modify accretion onto component SMBHs resulting in different features from that of
a single SMBH.

\subsubsection{Sound instability}
When rotation of the binary SMBHs is fast enough (namely separations are close enough),
a new instability arises depending on the competition of the orbital motion with the sound 
propagation (the cold part which has not been evaporated yet) and the evaporation. 
If both $\tau_{\rm evap}$ and $\tau_{\rm sound}$ are
longer than $P_{\rm orb}$, configuration of the mini-disks will be destroyed 
and evaporation of the disturbed disk be quenched. 
From $\tau_{\rm evap}=P_{\rm orb}$, we have the evaporating radius within one orbital period,
\begin{equation}\label{Eq:revap}
r_{\rm evap}=11.5\,\left(\alpha_{0.1}\dotm\right)^{0.21}\beta_{0.5}^{0.63}a_{2}^{0.32}
             (1+q)^{-0.11}f_{0.23}^{-0.21}.
\end{equation}
In principle, we should solve the non-linear equation for $r_{\rm evap}$ because of the
factor $f$. It turns out that $r_{\rm evap}\approx 11.5$ is a good approximate solution 
of Eqn.(\ref{Eq:revap}) and $f_{0.23}\approx 1$.

Moreover, when the sound timescale $\tau_{\rm sound}$ at $r_{\rm evap}$ is shorter than 
the orbital period $P_{\rm orb}$, the un-evaporated part of the inner region is suffering 
from the sound instability. In such a case, if a radial perturbation happens, for example, 
tidal interaction with its companion SMBH, the mini-disks will be highly twisted so that 
the disks will be completely deformed and cannot recover within one orbital period, 
destroying the disk configuration of accretion\footnote{Accretion flows can be transonic 
at some radii of a few $\Rg$ \citep{Muchotrzeb1981,Abramowicz1988}, but the radii are 
much smaller than the evaporation radius. The transonic effects can be neglected in the
sound instability.}. This 
is a new kind of instability, and is denoted as sound instability. In the meanwhile, 
evaporation is not able to proceed outward even for stationary hot corona.
With help of Eqn. (\ref{Eq:revap}), we have a critical separation from 
$\tau_{\rm sound}=P_{\rm orb}$,
\begin{equation}
a_{\rm crit}=75.4\,\alpha_{0.1}^{0.76}\beta_{0.5}^{2.25}\dotm^{-0.66}
             (1+q)^{0.33}f_{0.23}^{2.19},
\end{equation}
where $a_{\rm crit}=A_{\rm crit}/\Rg$. 

Detailed processes of the sound instability are highly non-linear. However the 
instability can not be suppressed once it is triggered. For example, the Coriolis 
force will efficiently stretched the mini-disk along the $\phi$-direction. Moreover,
the contraction of the binary separations will increase with orbital evolution 
and the $r-$ perturbation holds until merger. It is interesting to note that this critical
separation is independent of SMBH mass, but mildly dependent of accretion rates and sensitive
to the magnetic fields. 
For $q\sim 1$, we have $a_{\rm crit}\approx 95$. CB-SMBHs with separations less than 
$a_{\rm crit}$ are not allowed to have their own mini-disks. This new kind of 
instability purely arises from
the competition between the orbital motion and sound propagation. Without cold mini-disks, 
the binary SMBHs will radiate peculiar form of spectral energy distributions.

As to Case A and C, the sound instability is suppressed. 
Comparing $P_{\rm orb}$ with sound timescales of super-Eddington accretion, we find 
that it is always longer than $\tau_{\rm sound}$ (even for $r\sim a$), and
super-Eddington is stable under the orbital motion. It is then
expected that SEDs from super-Eddington accretion known as $F_{\nu}\propto \nu^{-1}$
holds \citep{Wang1999,Wang1999b} in the merging binary SMBHs. Component SMBHs of 
the merging binaries with super-Eddington accretion rates share one common broad-line 
region (BLR). They should show spectral features of strong optical \feii\, and weak 
\oiii\, lines in super-Eddington accreting AGNs \citep{Du2019}.

In a summary, mini-disks around each SMBH are undergoing the sound instability due 
to the fact that orbital motion is faster than the sound propagation when the binary 
separations are close enough. The mini-disks cannot exist due to the instability 
unless it has highly super-Eddington rates. The sound instability drives a conversion 
of cold mini-disks into hot gas following the orbital motion. 

We would like to point out validity of the present discussions. Corrections of timescales
in the co-moving frame of the fluid orbiting the mini-disks can be neglected since it is
at a level of ${\cal{O}}(10^{-2})$ for a region beyond $\sim 10^{2}\Rg$. We note that 
this can be justified by the pseudo-Newtonian potential of $\Psi=-G\bhm/(R-2\Rg)$ 
by \cite{Paczynski1980}. We note that magnetic field might stabilize the mini-disks 
when the fields are strong enough as shown in numerical simulations \citep{Cattorini2022}. 
The sound instability discussed in this paper could be weakened. The issues of mini-disks 
are highly non-linear and detailed analysis should be done from dynamical equations.

\section{Accretion onto binaries of SMBHs}
\subsection{Bondi accretion}
Gas remains in the cavity, and its density and temperature are determined by the energy 
supplied by the rotation energy of the binary. The orbital velocity of the primary SMBH 
with respective to the mass center is $\upsilon_{\bullet}/c=q(1+q)^{-1/2}a^{-1/2}$,
which is much faster than the disk sound speed. Shocks formed by the
fast orbital motion with a Mach number of 
${\cal M}\sim \upsilon_{\bullet}/c_{s}\sim 10^{3}$ will efficiently heat the
cavity gas. 
SMBHs gravitationally bound more gas at earlier stage of the interaction when the gas 
is not hot so that SMBHs have very large cross section colliding with cavity gas. 
Heating details are complicated, but the temperature is expected to be close to the 
virial temperature, namely,
$T\sim m_{\rm p}c^{2}/kr\approx 1.1\times 10^{11}r_{2}^{-1}\,${\rm K}. In such a hot 
plasma, the Coulomb coupling between proton and electrons are so inefficient 
that electron temperatures are much lower than that of protons \citep{Stepney1983}. 
Electrons keep temperatures $\sim 10^{9}\,$K
\citep[e.g.,][]{Shapiro1976,Rees1982} and Eqn. 40 \citep[in][]{Mahadevan1997}. 
We take $T=10^{9}\,T_{9}\,$K in this paper. Cavity gas is cooling through free-free 
emission with a timescale
\begin{equation}
\tau_{\rm ff}=5.7\times 10^{3}\,T^{1/2}n_{e}^{-1}\,{\rm yr},
\end{equation}
where $n_{e}$ is the number density of electrons in the cavity, and $T$ is the temperature 
of cavity gas. The cooling should be balanced by heating due to rotation of the binary,
this implies $\tau_{\rm ff}= \tau_{\rm gw}$, we have
\begin{equation}
n_{e}=6.0\times 10^{6}\,T_{9}^{1/2}a_{2}^{-4}\mu M_{8}^{-1}\,{\rm cm^{-3}}.
\end{equation}
SMBHs cannot have their own mini-disks with accretion rates of the standard model, 
however, they are still accreting in Bondi mode with a rate of \citep[e.g.,][]{Kato2008}
\begin{equation}
\dot{M}_{\rm Bon}=\frac{4\pi G^{2}\bhm^{2}n_{e}m_{\rm p}}{\left(c_{s}^{2}+\upsilon_{\bullet}^{2}\right)^{3/2}}
                 \approx \frac{4\pi G^{2}\bhm^{2}n_{e}m_{\rm p}}{\upsilon_{\bullet}^{3}},
\end{equation}
in the cavity context, 
where $m_{\rm p}$ is the proton mass. The sound speed of the cavity is poorly known, but
it cannot exceed the orbital velocity. Given the cavity density, we have dimensionless 
accretion rates of $\mathdotM_{\rm Bon}=\dot{M}_{\rm Bon}/{\dot{M}_{\rm Edd}}$
\begin{equation}\label{eq:dotm}
\mathdotM_{\rm Bon}=0.1\,q^{-3}(1+q)^{3/2}a_{2}^{3/2}n_{7}M_{8},
\end{equation}
where $n_{7}=n_{e}/10^{7}\,{\rm cm^{-3}}$. It should be noted that Eqn. 
(\ref{eq:dotm}) is only valid for the primary when $\upsilon_{\bullet}\gg c_{s}$.

We estimate the electron density of the hot cavity from balance between cooling and 
heating. Cavity gas is partially from evaporation and also from the twisted 
mini-disks. The Bondi spheres of component SMBHs greatly enhance the efficiency of 
heating gas in the cavity, but also significantly affects gas supply from the CBD.
Detailed transition of mini-disks into hot gas is of great interests as to AGN 
variabilities, and should be investigated by numerical simulations.
   
\subsection{Circumbinary disk}
A cavity of accretion disk will be formed through the tidal torque of the binary with 
accretion disk. We approximate the binary SMBHs as a single one in the circumbinary disk 
(CBD), the effective temperature is
\begin{equation}
T_{\rm eff}=\left(\frac{3G\bhm \dot{M}}{8\pi\sigma_{\rm SB}R^{3}}\right)^{1/4}
           =1.9\times 10^{4}(\mathdotM/M_{8})^{1/4}r_{2}^{-3/4}\,{\rm K},
\end{equation}
where $r_{2}=R/10^{2}\Rg$ is the radius of the CBD to the mass center of the binary, and
the boundary factor $f\approx 1$ at $r=10^{2}$. The CBD is radiating photons with energies 
of 
\begin{equation}\label{Eq:epsilon}
\epsilon_{\rm CBD}\approx 4.5\,(\mathdotM/M_{8})^{1/4}r_{2}^{-3/4}\,{\rm eV},
\end{equation}
which is significantly lower than the ionization energy of hydrogen 
($\epsilon_{\rm H}=13.6\,{\rm eV}$) and first level of 
magnesium atoms ($\epsilon_{\rm Mg^{+}}=7.6\,{\rm eV}$). 

\begin{figure*}
\centering
\includegraphics[width=0.75\textwidth]{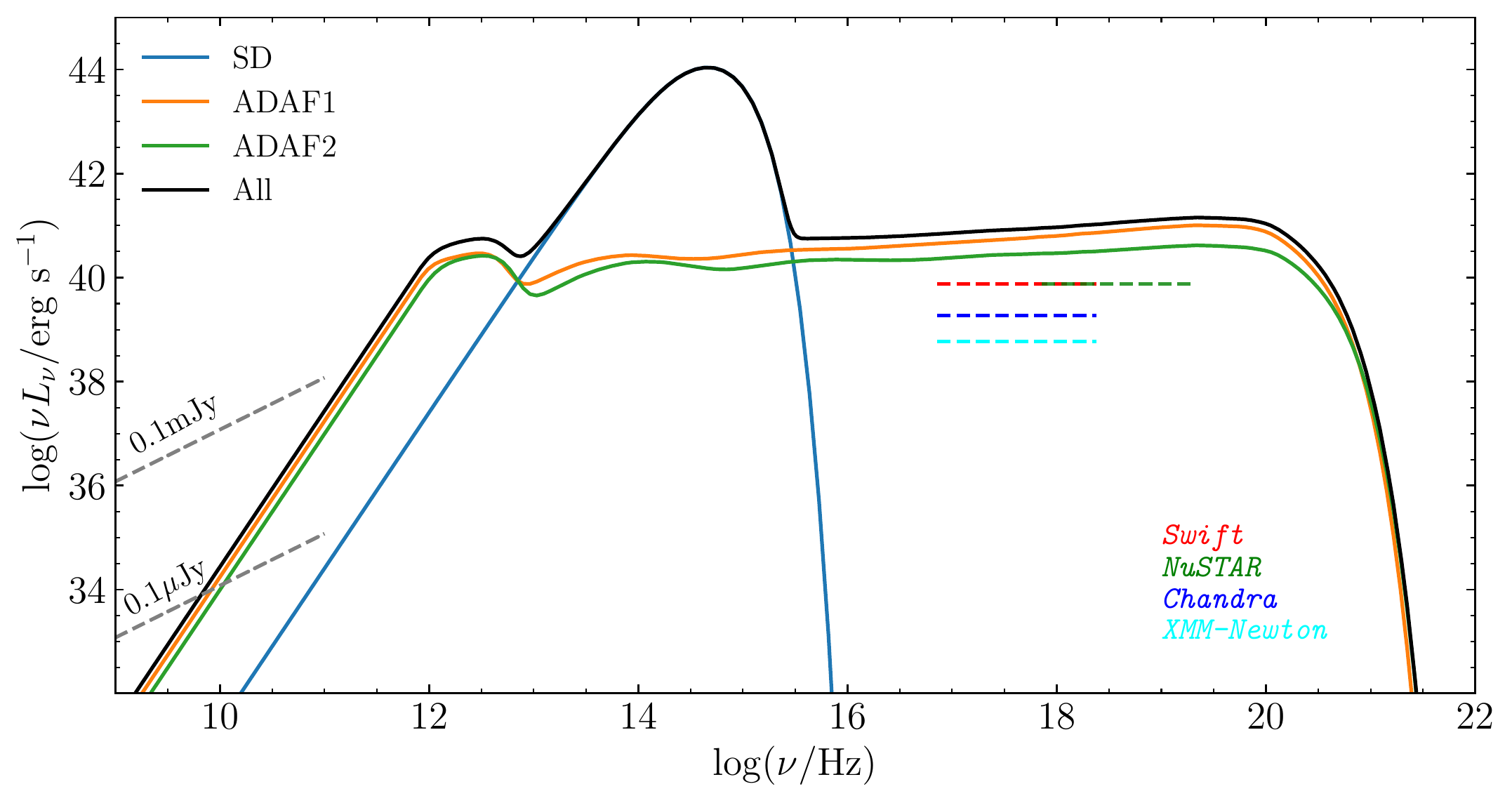}
\caption{Characterized shapes of spectral energy distributions 
of the binary system: 1) two-ADAF 
emissions: yellow line for $(\bhm/\sunm,\mathdotM)=(10^{8},0.1)$ and green for 
$(10^{7},0.07)$; and 2) CDB emissions (blue line for $\mathdotM=1$) from standard disk (SD). 
We assume that SMBHs are Schwartzschild black holes, outer and inner boundary radii are 
($50,6)\,\Rg$. Radio emissions from ADAFs are expected to undergoing periodic variabilities. 
Optical emissions are similar to normal AGNs, but they are radio-quiet and X-rays are weak. 
Detection levels of observations from radio to X-rays are marked, showing feasible detections 
in X-rays.
}
\vglue 0.3cm
\label{fig:sed}
\end{figure*}

In this paper, we focus on the case of binary SMBHs with separations less than 
$A_{\rm crit}$. The two SMBHs share a common broad-line region (BLR), and hence 
the BLR follows the well-known $R-L$ relation \citep{Kaspi2000,Bentz2013,Du2019}.
For a CBD with $\mathdotM\sim 1$, Eqn. (\ref{Eq:epsilon}) indicates the lack 
of ionizing photons in BLR. The merging binaries of SMBHs are expected to not 
have significant emission lines such as Balmer lines and \mgii$\lambda2798$\AA\ 
though they have normal optical continuum as AGNs. However, \mgii\, line appears
when $\mathdotM\gtrsim 9$ (slightly super-Edddington; see \citealt{Du2019}) but 
it could be weak. When $\mathdotM\gtrsim 90$, Balmer lines just appear and \mgii\ line 
could be normal, but high ionization line \civ$\lambda1549$\AA\, 
($\epsilon_{\rm C^{3+}}=47.9\,$eV)  will never appears unless $\mathdotM\gtrsim 10^{4}$ 
much exceeding the known accretion rates of SEAMBHs \citep{Du2019}. Generally, these 
merging binaries appear as optically bright nuclei without strong broad emission lines.


\section{Observational signatures}
\subsection{SEDs}
Emergent spectra of the merging binaries are composed of two parts: 1) ADAF emissions 
from the primary and secondary SMBHs and 2) CBD emissions. Emissions 
from the cavity can be neglected if compared with Bondi accretion\footnote{According 
to self-similar solution of ADAF \citep[see their Eq.\,12.5 in][]{Narayan1995}, we have ADAF 
gas density of $\sim 10^{8}\,{\rm cm ^{-3}}$ at $r=10^{2}$ for $\mathdotM=0.1$. Since the 
cavity density of hot gas is about $3\%$ of the ADAF, cavity gas emission
is only of $\sim 10^{-3}$ of the ADAF from free-free emissions of hot plasma.}. 
We apply the ADAF model 
to merging SMBHs for their emergent spectra formulated by \cite{Manmoto2000} 
\citep[see also][]{Li2009}. Bolometric luminosity of ADAFs is about
$L_{\rm ADAF}\approx 3.0\times 10^{43}\,\alpha_{0.3}^{-2}M_{8}\mathdotM_{0.1}^{2}\,\ergs
\propto \mathdotM^{2}$
\citep[see Eqn.\,49 in][]{Mahadevan1997}, where $\mathdotM_{0.1}=\mathdotM/0.1$.
Figure \ref{fig:sed} shows the characteristic emergent spectra.
Emissions from the SMBH Bondi accretion span from radio to X-rays, but its powers 
are significantly lower than the CBD. For one target at a distance of $100\,$Mpc, radio
fluxes received by the observers are about $F_{100\rm GHz}\approx 2.6\,\mu$Jy at $\nu=100$GHz,
which is significantly higher than sensitivity of SKA ($0.1\,\mu$Jy), but much lower than 
the current VLA sensitivity. In the meanwhile, we also mark the detectable level of 
sensitivity of X-ray missions for future detections ({\it Chandra}, {\it NuSTAR},
{\it XMM-Newton} and {\it Swift}) at 100\,Mpc. It is clear that the current missions
are able to detect the periodic variations. Some weak-line quasars have similar SEDs
to the predicted shape in \cite{Wu2012} and \cite{Luo2015} samples ({\it GALEX} data 
were used) and X-shooter sample \citep{Plotkin2015}, but the UV deficit should be 
necessarily investigated by more observations. Moreover, Doppler boosting can be
tested with the multiwavelength continuum.

Since optical emissions are from the CBD in standard regime, it is expected that radio emissions
are much fainter than optical and they will show radio-weak or radio-quiet properties.
Comparing with optical emissions, X-rays are much fainter showing X-ray weak. Radio emissions
depend on the outer boundary radius, where we take it to be $50\Rg$. Future detection at 
$\sim 10-100\,$GHz will reveal the radio periodical variations if merging binaries exist.

For super-Eddington accretion onto each components of the binaries, its optical to UV
continuum are characterized by the known spectra of $L_{\nu}\propto \nu^{-1}$. This results 
from the effective temperature distribution of $T_{\rm eff}\propto r^{-1/2}$ due to photon 
trapping effects \citep{Wang1999,Wang1999b}.

SEDs of periodical AGNs and quasars \citep{Graham2015a,Charisi2016,Liu2019} 
from radio to UV (GALEX data) are found to be similar to that of normal AGNs and quasars, 
and the periodicity of the sample identified from $\gtrsim 1.5\,$cycle light curves may be 
false positive \citep{Guo2020}. On the other hand, periodicity of AGNs and quasars is needed
to be confirmed with longer baselines.

\begin{figure*}
\centering
\includegraphics[width=0.95\textwidth]{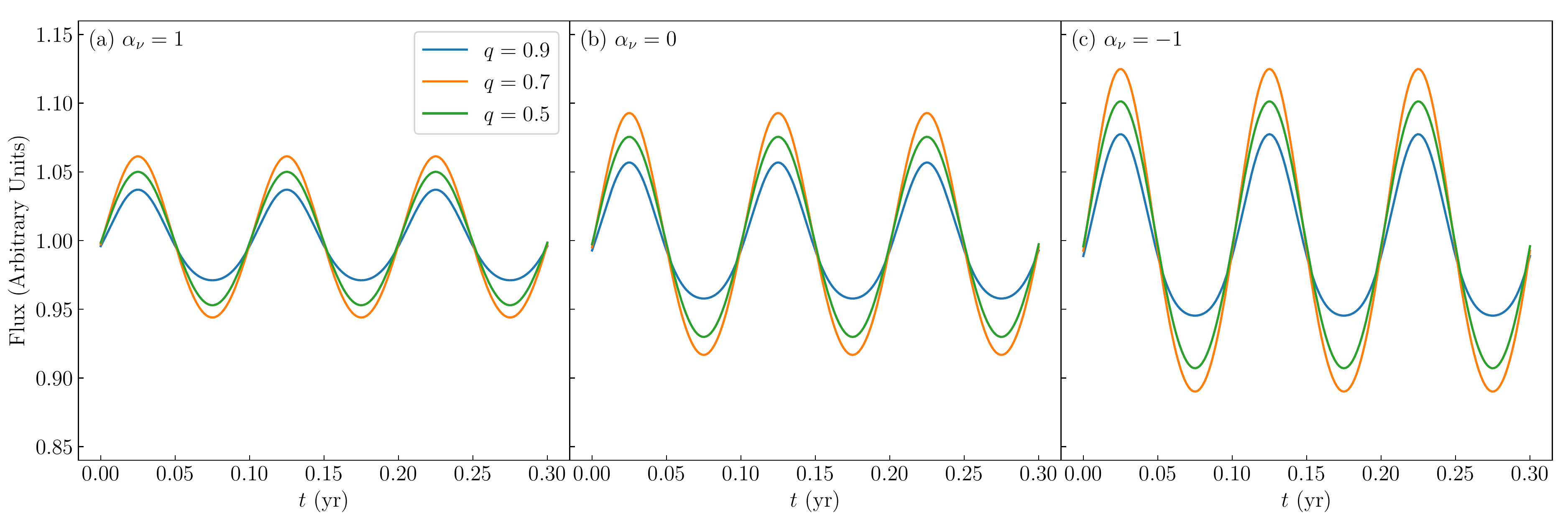}
\caption{Radio light curves of binary SMBHs with ADAF emissions modulated by
Doppler boosting. We take the primary SMBH of $\bhm=10^{8}\sunm$, and a period of $0.1\,$yr. 
An inclination angle of $30^{\circ}$ is assumed. Here we plot three indexes for different 
bands: 1) cm band ($\alpha_{\nu}\approx 1$); 2) mm band ($\alpha_{\nu}\approx 0$) and 3) 
sub-milimter ($\alpha_{\nu}\approx-1$). Though the periods of different bands are same but 
the amplitudes are dependent of wavelengths. If X-ray light curves are obtained through 
monitoring campaigns, the dependence of Doppler boosting on wavelength can be tested.}
\label{fig:LC}
\end{figure*}

\subsection{Doppler boosting}\label{sec:Doppler}
\cite{DOrazio2015} suggested that orbital motion of binary SMBHs causes periodical
variations due to Doppler boosting effects if the separations are small enough.
Supposing that the orbital plane ($XOY$) has an inclination angle of $i_{\rm o}$ with 
respect to observer's line of sight (in $XOZ$ plane), the direction of observers 
is $\vec{n}_{\rm o}=(\sin i_{\rm o},0,\cos i_{\rm o})$ and the SMBH are 
$\vec{n}_{\bullet}=(\cos\phi,\sin\phi,0)$, where $\phi=2\pi t/P_{\rm orb}$ is 
the phase angle. The angle between SMBH motion direction and the observers
is given by
$\cos\theta=\vec{n}_{\rm o}\cdot\vec{n}_{\bullet}=\sin i_{\rm o}\cos\phi$.
The Doppler factor is given by
\begin{equation}
\calD=\frac{1}{\gamma(1-\beta\cos\theta)},
\end{equation}
where $\gamma=1/\sqrt{1-\beta^{2}}$ is the Lorentzian factor of SMBHs and $\beta$ is 
the velocity in units of light speed. Considering the invariance of $F_{\nu}/\nu^{3}$, 
we have the total fluxes from the two SMBHs
\begin{equation}
F_{\rm tot}=\sum_{i=1}^{2}\frac{F_{i}}{\left[\gamma_{i}
                      (1-\beta_{i}\cos\theta_{i})\right]^{3-\alpha_{\nu}}},
\end{equation}
where $F_{i}$ is the spectral flux from the $i$-th ADAF, 
$\beta_{1}=q(1+q)^{-1/2}a^{-1/2}$ and $\beta_{2}=(1+q)^{-1/2}a^{-1/2}$ are
velocities of the primary and the secondary SMBHs around the binary mass center, 
respectively, their projected angles are
$\cos\theta_{1}=\sin i_{\rm o}\cos(2\pi t/P_{\rm orb})$ and 
$\cos\theta_{2}=\sin i_{\rm o}\cos(2\pi t/P_{\rm orb}+\pi)$, and $\alpha_{\nu}$ is the 
spectral index of ADAFs. Path differences of photons from the primary and the
secondary SMBHs are neglected since $A/c\ll P_{\rm orb}$.

According to the scaling law of radio luminosity 
$L_{\rm rad}\propto \mathdotM^{3/2}$ from ADAFs \citep[see Eqn. 24 in][]{Mahadevan1997}, 
we have $F_{1}\propto q^{-9/2}(1+q)^{3/4}$ and $F_{2}\propto q^{11/2}(1+q^{-1})^{3/4}$. 
For radio spectra of ADAFs, we take $\alpha_{\nu}\approx 1.5$ in our calculations.
Radio fluxes from ADAFs are very sensitive to mass ratios due to the 
Bondi accretion as well as inclination angle. In Figure \ref{fig:LC}, 
we show the periodic light curves of ADAF radio emissions. It is expected that X-rays 
should have the same periods with radio, but variation amplitudes are different from
radio because of different spectral indices from radio bands.

\subsection{Warped disks}
Radio emissions should be periodic, and optical emissions are also periodic if the 
CBD is warped with a tilt angle ($\beta_{\rm t}$). The processing CBD will show periodic 
variations due to emitting area varying with time projected to observers 
\citep[e.g.,][]{Martin2007}, namely, we are receiving photons from periodically-changing 
surface of the warped disk. The CBD precession driven by tidal torque of the binary with 
a circular orbit will show a modulations of received fluxes with a period given 
by \cite{Hayasaki2015}
\begin{equation}
\tau_{\rm pre}=\frac{4(1+q)^{2}}{3\pi q\cos \beta_{\rm t}}\left(\frac{R}{A}\right)^{7/2}
               P_{\rm orb}\approx 19.2\,P_{\rm orb},
\end{equation}
for $\cos \beta_{\rm t}=0.5$ and $q=1$, where we take $R/A=2$. This implies that the 
optical periods are ten times the radio periods. Target selections of radio periodic
AGNs could be favored from those with optical periods of a few years.

\subsection{Period changes}
Period changes due to GW (see Eqn.\ref{eq:rates}) can be exactly measured by radio and X-ray 
campaigns as well as X-ray light curves. How to measure the orbital parameters of the merging
binary SMBHs is a challenging task. Fortunately, SARM can accurately measure the total mass 
of binary SMBHs with the weak lines, like for 3C\,273 
\citep[the best measurements with accuracy of about 15\% in][]{Wang2020c,Li2022}. 
With orbital 
parameters, we can estimate the expected period changes and compare them with observations 
($\sim 0.3\,{\rm day\,yr^{-1}}$) from 
Eqn. (\ref{eq:rates}) provided radio periods are accurately measured. Such a test is feasible
for the merging binary SMBHs. This way is exactly same with that of \cite{Hulse1975} for binary 
pulsars in radio. Moreover, if future PTA becomes efficient enough to test individual 
merging binary SMBHs, low frequency GW can be observationally tested for gravity theory.

\subsection{Relativistic jets}
As one of well-known properties, relativistic jets/outflows can be naturally produced from 
ADAFs of SMBHs \citep{Narayan1994,Blandford2019}. It is thus to expect 
jets/outflows from SMBHs inside the hot cavity. The jet powers from extreme Kerr SMBHs
can be simply estimated by $L_{\rm jet}\approx \eta_{\rm BZ}\dot{M} c^{2}=
2.8\times 10^{43} \eta_{0.02}\mathdotM_{0.1} M_{8}\,\ergs$ \citep{Armitage1999},
where $\eta_{0.02}=\eta_{\rm BZ}/0.02$ is the radiative efficiency of the 
Blandford-Znajek (BZ) mechanism. This BZ process 
is much more efficient than ADAF emissions ($L_{\rm ADAF}\ll L_{\rm jet}$) 
if $\mathdotM\ll 0.1$. 

If the component SMBHs (extreme Kerr ones) 
are generating their own jets independently, double two-pair jets will be produced from 
the binary SMBH system \citep[e.g.][]{Paschalidis2021}, even there is the third jet 
produced by pumping rotating energy after the binary SMBHs merge. In the 
presence of relativistic jets, observers receive two kinds of radio components: 1) binary 
ADAFs and 2) double two-pair jets. Even the jets are misaligned to the observers, radio 
emissions are still quite strong as shown by $L_{\rm jet}$. The light curves will be 
complicated and depend on orientations of jets ($\Theta_{\rm o}$) and their Lorentzian 
factor ($\Gamma$). The observed luminosity from the jets is 
$L_{\rm obs,\nu}={\cal D}_{\rm jet}^{3-\alpha_{\nu}}L_{\rm jet,\nu}$, where 
${\cal D}_{\rm jet}$ is the Doppler factor, and ${\cal D}_{\rm jet}\sim 10$ is common
for jets but depends on $\Theta_{\rm o}$. Jet's apparent luminosity is much brighter than
ADAF itself. When jets are spiraling outward, and the relativistic Doppler factor is 
function of time, showing periodic variations.  This is similar to the jet case in 
SS\,433 \citep{Hjellming1981}, whose periodic variations are caused by orbital and 
precession of jets. We will separately discuss this issue in a future paper.

\section{Searching for merging binaries}
Merging binary SMBHs are characterized by several features: 1) weak broad emission lines
(\mgii\, and H$\beta$); 2) continuum with UV deficit (unless those candidates with 
super-Eddington rates); 3) plausibly periodic variations in radio, optical and X-ray 
bands, but the two periods could be very different.  

It remains open to estimate SMBH masses as a big issue of estimating separations of 
SMBH binaries. Generally SMBH masses can be estimated by Magorrian relation or $\bhm-\sigma$ 
relation for radio galaxies or weakly active galactic nuclei \citep[e.g.,][]{Kormendy2013}. 
If the hosts of SMBH binaries are normal AGNs, we can employ the technique of reverberation 
mapping not only for SMBH mass but also orbital parameters including ellipticity, mass ratios 
and semi-major axis \citep{Wang2018,Songsheng2020,Kovacevic2020b}. This needs high spectral 
resolution spectrograph installed on 8m telescope \citep{Songsheng2020}. Joint analysis of 
GRAVITY/VLTI and RM data will be more robust for orbital parameters.

\subsection{Weak line quasars and optical periodicity}
Weak line quasars (WLQs), which are usually defined by EW(\civ) less than $10\,$\AA, 
have been discovered by SDSS \citep{Fan1999,Fan2006,Anderson2001,Collinge2005,Diamond2009}, 
however, nature of their multiple SEDs remains open so far 
\citep{Wu2012,Plotkin2015,Luo2015,Ni2022}. 
They are either radio-quiet or radio-loud \citep{Plotkin2010}, 
and about half of WLQs are X-ray weak \citep{Luo2015,Ni2022}, however, the UV continuum 
is uncertain somehow. Some WLQs seem to have SEDs similar to our prediction in this work
(see X-shooter spectrum in some WLQs in \citealt{Plotkin2015}). It will be interesting to 
test if these WLQs have periodic variations at optical bands.

\cite{Guo2019} investigate UV continuum of \cite{Graham2015a} sample to check their UV 
deficit, however, they have not found this deficit from this sample. On the other hand, 
UV deficit could be caused by dusty reddening/extinction, for example, Mrk 231
\citep{Yan2015,Leighly2016}. On the one hand, periodicity  of \cite{Graham2015a},
\cite{Charisi2016} and \cite{Liu2019} samples should be tested by extending baseline 
of light curves.  Dependence of Doppler boosting 
effects on different wavebands has been checked for some objects \citep{Charisi2016,Chen2021}. 
Moreover, it should be mentioned that only periodicity of optical variations is not enough 
to justify binary SMBHs. It is far away to draw conclusions as to binarity of SMBHs in 
the current sample. Large Synoptic Survey Telescope (LSST) is highly expected to show 
high quality light curves of AGNs to search for candidates of merging SMBHs in near future. 


\subsection{Radio periodic AGNs}
To search for the binaries, we should monitor AGNs for periodicity.
Time-domain survey of VLASS has been designed by \cite{Lacy2020} (PKS 2131-021 is one of 
the targets of VLASS). This lends opportunity of searching for radio periodic light curves 
of the candidates of merging binary SMBHs. With great progress of optical periodic AGNs, 
we should pay more attention to them in radio (also in X-rays).

We would mention the possibility that the binary ADAFs inside the cavity could
produce relativistic jets. If jets are in alignment with observer's line of sight,
we will see a radio-loud AGNs. 
\cite{King2013} reported a radio light curve with a period of $120-150$\,days in the 
blazar J1359+4011, which is very different from other blazars with periodical variations. 
This was explained by disk oscillations similar to that in X-ray binary because this 
period is much shorter than the known physics of jet/disk precession \citep{King2013}. 
Alternatively, we might explain this from the orbital Doppler boosting of merging 
binary SMBHs. 
Recently, \cite{Zhang2022} report double periods of $345\,$day and $386\,$day 
in NGC\,1275 (3C\,84), providing indirect evidence for hypothesis of close binary SMBHs. 
Recently, radio periodicity of PKS 2131-021 is found as a candidate of binary 
SMBHs \citep{ONeill2022}.

\subsection{Detection probabilities}
Considering AGN lifetimes ($\tau_{\rm AGN}$), we may detect the merging binary SMBHs 
with a probability of $p\approx \tau_{\rm gw}/\tau_{\rm AGN}\sim 3\times 10^{-6}$ 
if $\tau_{\rm AGN}=10^{7}\,{\rm yr}$ is taken (which is the radial timescale
of accretion flows at its outer boundary). For the current sample of million AGNs and 
quasars\footnote{https://heasarc.gsfc.nasa.gov/W3Browse/all/milliquas.html}, we may have 
a few cases. However, we should note that $\tau_{\rm gw}$ is very sensitive to the 
separations of binaries and $\tau_{\rm AGN}$ is highly uncertain.
If we take $\tau_{\rm AGN}\sim 10^{5}\,$yr \citep{Schawinski2015}, we have 
$p\sim 3\times 10^{-4}$ implying $\sim 300$ AGNs as merging binary SMBHs in the Universe. 
This is a quite large number. Moreover, this crude estimation could be improved by including 
evolution of orbits.

In summary, searching for candidates from WLQs with periodical variations is feasible in 
radio and X-ray bands, in particular optical (ZTF and LSST). Moreover, UV deficit is 
explored by the X-shooter and HST UV 
observations which is also a key to explain nature of WLQs.

\section{Summary}
Merging binary supermassive black holes are completely elusive in observations because
accretion onto the binaries is insufficiently understood. In this paper, we investigate 
fates of the mini-disks with three regimes of accretion rates from low to super-Eddington
status. In the regime of standard 
accretion disks, the innermost regions of the mini-disks are evaporated through thermal 
conduction of hot corona, and the mini-disks are truncated. We find 
a new instability of the mini-disks when the binary SMBHs are close enough.
In such a context, the sound propagation timescale is longer than the orbital period
of the binaries so that the disks are highly twisted and destroyed. For binary SMBHs
of $\sim 10^{8}\,\sunm$ with typical orbital values, the critical separation is about 
$\sim 10^{2}$ gravitational radius (orbital periods are about $0.1\,$yr). For low accretion 
and super-Eddington regimes, the sound instability disappears. 

Heated by the orbital motion of the binaries, cavity is then full of hot gas and 
SMBHs are accreting through Bondi mode unless binaries with extreme super-Eddington 
accretion rates. Emissions from the Bondi accretion are modulated by the orbital 
motion and Doppler boosting effects. For radio-quiet AGNs, radio monitoring campaigns
are expected to detect periodic variations at these bands. If normal 
galaxies contain binary SMBHs, they will show similar periodical variations, in 
particular, radio and X-ray campaigns of monitoring candidates are suggested because 
of less contaminations from host galaxies.

For binary SMBHs with intermediate accretion rates, the UV deficit as a continuum 
characteristic appears and AGNs thus show weak broad emission lines. The circumbinary 
disk is radiating in optical band whereas Bondi accretion of SMBHs in the cavity 
is generating multiwaveband continuum from radio to hard X-rays. Moreover, Balmer 
lines disappear owing to the significant lack of ultraviolet photons. However, for 
binaries with super-Eddington rates, spectral energy distributions keep the form
of $F_{\nu}\propto \nu^{-1}$ without UV deficit. Emission line features should be
similar to those SEAMBHs (strong optical \feii\, and weak \oiii\, lines).
There is an interesting case of warped disks showing periodical 
variations in optical bands with periods of ten times the radio periods if the 
circumbinary disks are misaligned with the orbital plane. 

Searching for merging binary SMBHs is very challenging since they could occupy a very tiny
fraction ($\sim 10^{-6}-10^{-3}$) of the current AGN sample.
Candidates of merging SMBHs are expected to show observational signatures characterized 
by ultraviolet deficit with/without weak emission lines as well as X-ray weak emissions 
(and potentially with optical periods). For extreme Kerr SMBHs in merging binaries, 
they will generate relativistic jets. In this case, calculations of light curves will 
be complicated and beyond the scope of this paper. Radio and X-ray periodic variations 
are the key diagnostic of the binaries, current VLA can focus on a systematic search from 
radio-loud AGNs and future SKA from radio-quiet ones \citep[e.g.,][]{Bignall2015}. 

\section*{Acknowledgements}
JMW is grateful to L. C. Ho for interesting discussions, and to E. Qiao for evaporation
processes. This research is supported by grant 2016YFA0400700 from the Ministry of Science 
and Technology of China, by NSFC-11991050,
-11773029, -11833008, -11690024, -11721303, -11991052
-11573026, -11873048, -11703077, -11503026, and by the CAS Key Research
Program (KJZDEW-M06) and by the Key Research Program of Frontier Sciences,
CAS, grant QYZDJ-SSW-SLH007.

\section*{DATA AVAILABILITY}
The data underlying this article will be shared on reasonable request to the 
corresponding author.

\appendix
\section{Solutions of standard accretion disk models}
Mini-disks as accretion flows of component SMBHs mainly involve the inner region, where 
radiation pressure and scattering dominate. We employ the solutions of standard accretion 
disk models \citep{Shakura1973}, radial velocity, temperature, surface density, mass density 
given by
\begin{equation}
\upsilon_{r}=1.4\times 10^{6}\,\alpha_{0.1}\mathdotM^{2}r_{1}^{-5/2}f\,{\rm cm\,s^{-1}},
\end{equation}
\begin{equation}
T_{c}=4.8\times 10^{5}\,(\alpha_{0.1} M_{8})^{-1/4}r_{1}^{-3/8}\,{\rm K},
\end{equation}
\begin{equation}\label{Eq:surface}
\Sigma=1.1\times 10^{4}\,\left(\alpha_{0.1}\dotm\right)^{-1}r_{1}^{3/2}f^{-1}\,{\rm g\,cm^{-2}},
\end{equation}
and
\begin{equation}
\rho=1.0\times 10^{-10}\,\left(\alpha_{0.1}M_{8}\right)^{-1}\dot{m}^{-2}r_{1}^{3/2}f^{-2}\,
         {\rm g\,cm^{-3}},
\end{equation}
respectively, in the inner region \citep[e.g.,][]{Kato2008}. 

\nolinenumbers


\begin{thebibliography}{}
%
\bibitem[Abramowicz et al.(1988)]{Abramowicz1988}
Abramowicz, M., Czerny, B., Lasota, J.-P. \& Szuszkiewicz, E. 1988, \apj, 332, 646


\bibitem[Anderson et al.(2001)]{Anderson2001}
Anderson, S. F., Fan, X., Richards, G. T., et al. 2001, \aj, 122, 503

\bibitem[Armengol et al.(2021)]{Armengol2021}
Armengol, F. G. L., Combi, L., Campanelli, M., et al. 2021, \apj, 913, 16,
 
\bibitem[Armitage \& Natarajan(1999)]{Armitage1999}
Armitage, P. J. \& Natarajan, P. 1999, \apjl, 523, L7
 
\bibitem[Artymowicz \& Lubow(1994)]{Artymowicz1994}  
Artymowicz, P., \& Lubow, S. H. 1994, ApJ, 421, 651 

\bibitem[Begelman et al.(1980)]{Begelman1980}
Begelman, M. C., Blandford, R. D. \& Rees, M. J. 1980, \nat, 287, 307

\bibitem[Bentz et al.(2013)]{Bentz2013}
Bentz, M. C., Denney, K. D., Grier, C. J., et al. 2013, \apj, 767, 149

\bibitem[Bignall et al.(2015)]{Bignall2015}
Bignall, H. E., Croft, S., Hovatta, T. et al. Proceedings of Advancing 
Astrophysics with the Square Kilometre Array. arXiv:1501.04627 

\bibitem[Blanchet(2014)]{Blanchet2014}
Blanchet, L. 2014, Living Rev. Relativity, 17, 2

\bibitem[Blandford et al.(2019)]{Blandford2019}
Blandford, R., Meier, D. \& Readhead, A. 2019, \araa, 57, 467

\bibitem[Bogdanovi\'c et al.(2021)]{Bogdanovic2021}
Bogdanovic, T., Miller, M. C. \& Blecha, L. 2021,  arXiv:2109.03262 

\bibitem[Bowen et al.(2018)]{Bowen2018}
Bowen, D. B., Mewes, V., Campanelli, M., et al. 2018, \apjl, 853, L17
 
\bibitem[Bowen et al.(2019)]{Bowen2019}
Bowen, D. B., Mewes, V., Noble, S. C., et al. 2019, \apj, 879, 76 

\bibitem[Brotherton et al.(2020)]{Brotherton2020}
 Brotherton, M. S., Du, Pu, Xiao, M. et al. 2020, \apj, 905, 77
 
\bibitem[Cattorini et al.(2021)]{Cattorini2021}
Cattorini, F., Giacomazzo, B., Haardt, F., \& Colpi, M. 2021, PhRvD, 103, 103022



\bibitem[Charisi et al.(2016)]{Charisi2016}
 Charisi, M., Bartos, I., Haiman, Z., et al., 2016, \mnras, 463, 2145

\bibitem[Charisi et al.(2022)]{Charisi2022}
Charisi, M., Taylor, S. R., Runnoe, J., Bogdanovic, T., \& Trump, J. R. 2022, \mnras, 510, 5929

\bibitem[Chen et al.(2020)]{Chen2020}
 Chen, Y.-C., Liu, X., Liao, W.-T. et al. 2022, \mnras, 499, 2245
 
\bibitem[Chen et al.(2021)]{Chen2021}
Chen, Y.-C., Liu, X., Liao, W.-T. \& Guo, H. 2021, \mnras, 507, 4638 

\bibitem[Collinge et al.(2005)]{Collinge2005}
Collinge, M. J., Strauss, M. A., Hall, P. B. et al. 2005, \aj, 129, 2542

\bibitem[Combi et al.(2021)]{Combi2021}
Combi, L., Armengol, F. G. L., Campanelli, M. et al. 2021, \apj, arXiv:2109.01307

\bibitem[Cattorini et al.(2022)]{Cattorini2022}
Cattorini, F., Maggioni, S., Giacomazzo, B. et al. 2022, \apj, 930, L1


\bibitem[Czerny(2019)]{Czerny2019}
Czerny, B. 2019, Universe, 5, 131

\bibitem[Czerny et al.(2004)]{Czerny2004}
Czerny, B., R\'oza\'nska, A., \& Kuraszkiewicz, J. 2004, \aap, 428, 39 

\bibitem[Decarli et al.(2013)]{Decarli2013}
Decarli, R., Dotti, M., Fumagalli, M. et al. 2013, \mnras, 433, 1492

\bibitem[De Rosa et al.(2019)]{DeRosa2019}
De Rosa, A., Vignali, C., Bogdanovi\'c, T., et al. 2019, New A Rev., 86, 101525

\bibitem[Diamond-Stanic et al.(2009)]{Diamond2009}
Diamond-Stanic, A. M., Fan, X., Brandt, W. N., et al. 2009, \apj,  699, 782

\bibitem[Di Matteo et al.(2001)]{DiMatteo2001}
Di Matteo, T., Johnstone, R. M., Allen, S. W. \& Fabian, A. C. 2001, \apjl, 550, L19

\bibitem[Doan et al.(2020)]{Doan2020}
Doan, A., Eracleous, M., Runnoe, J. C., et al. 2020, \mnras, 491, 1104

\bibitem[D'Orazio et al.(2015)]{DOrazio2015}
 D'Orazio, D. J., Haiman, Z., Schiminovich, D. 2015, \nat, 525, 351

\bibitem[Dotti et al.(2022)]{Dotti2022}
Dotti, M., Bonetti, M., D'Orazio, Daniel J., et al. 2022, \mnras, 509, 212


\bibitem[Du et al.(2014)]{Du2014}
Hu, C., Lu, K.-X., et al. 2014, \apj, 782, 45

\bibitem[Du et al.(2015)]{Du2015}
Du, P., Hu, C., Lu, K.-X., et al. 2015, \apj, 806, 22

\bibitem[Du et al.(2018)]{Du2018}
 Du, P., Brotherton, M. S., Wang, K., et al. 2018, \apj,  869, 142
 
\bibitem[Du \& Wang(2019)]{Du2019} 
Du, P. \& Wang, J.-M. 2019, \apj, 886, 42

\bibitem[Ebisuzaki et al.(1991)]{Ebisuzaki1991}
Ebisuzaki, T., Makino, J., \& Okumura, S. K. 1991, \nat, 354, 212

\bibitem[Eggleton(1983)]{Eggleton1983}
Eggleton, P. P. 1983, \apj, 268, 368

\bibitem[Fan et al.(1999)]{Fan1999}  
Fan, X., Strauss, M. A., Gunn, J. E., et al. 1999, \apjl, 526, L57 

\bibitem[Fan et al.(2006)]{Fan2006}
Fan, X., Strauss, M. A., Richards, G. T., et al. 2006, \aj, 131, 1203

\bibitem[Farris et al.(2011)]{Farris2011}
Farris, B. D., Liu, Y. T., \& Shapiro, S. L. 2011, PhRvD, 84, 024024


\bibitem[Foord et al.(2021)]{Foord2021}
Foord, A., Liu, X., G\"ultekin, K. et al. 2021, arXiv2110:02982


\bibitem[Franceschini et al.(1998)]{Franceschini1998}
Franceschini, A., Vercellone, S. \& Fabian, A. C. 1998, \mnras, 297, 817

\bibitem[Giacomazzo et al.(2012)]{Giacomazzo2012}
Giacomazzo, B., Baker, J. G., Miller, M. C., Reynolds, C. S., \& van Meter, J. R. 
2012, \apjl, 752, L15

\bibitem[Gold et al.(2014)]{Gold2014}
Gold, R., Paschalidis, V., Ruiz, M., et al. 2014, PhRvD, 90, 104030

\bibitem[Graham et al.(2015a)]{Graham2015a} 
Graham M. J. et al., 2015a, \nat, 518, 74

\bibitem[Graham et al.(2015b)]{Graham2015b}
Graham, M. J., Djorgovski, S. G., Stern, D. et al. 2015b, \mnras, 453, 1562

\bibitem[GC(2017)]{GC2017}
Gravity Collaboration; Abuter, R.; Accardo, M. et al. 2017, \aap, 602, A94 (GC2017)

\bibitem[GC(2018)]{GC2018}
Gravity Collaboration; Sturm, E.; Dexter, J. et al. 2018, \nat, 563, 657 (GC2018)

\bibitem[GC(2020)]{GC2020}
Gravity Collaboration; Amorim, A.; Baub\"ock, M. et al. 2020, \aap, 643, 154 (GC2020)

\bibitem[GC(2021)]{GC2021}
Gravity Collaboration; Amorim, A.; Baub\"ock, M. et al. 2021, \aap, 648, 117 (GC2021)

\bibitem[Grossov\'a et al.(2022)]{Grossova2022}
Grossov\'a, R., Werner, N., Massaro, F. et al. 2022, \apjs, 258, 30

\bibitem[Guo et al.(2019)]{Guo2019}
Guo, H., Liu, X., Shen, Y., et al. 2019, \mnras, 482, 3288

\bibitem[Guo et al.(2020)]{Guo2020}
Guo, H., Liu, X., Zafar, T. \* Liu, W.-T. 2020, \mnras, 492, 2910

\bibitem[G\"ultekin \& Miller(2012)]{Gultekin2012}
G\"ultekin, K. \& Miller, J. M. 2012, \apj, 761, 90

\bibitem[Guti\'errez et al.(2022)]{Gutierrez2022}
Guti\'errez, E. M., Combi, L.,  Noble, S. C. et al. 2022, arXiv:2112.09773

\bibitem[Haardt \& Maraschi(1991)]{Haardt1991}
Haardt, F. \&  Maraschi, L. 1991, ApJ, 380, L51

\bibitem[Haiman et al.(2009)]{Haiman2009}
Haiman, Z.,  Kocsis, B. \&  Menou, K. 2009, \apj, 700, 1952

\bibitem[Hayasaki et al.(2015)]{Hayasaki2015}
Hayasaki, K., Sohn, B. W., Okazaki, A. T. et al. 2015, JCAP, 7, 5

\bibitem[Hjellming \& Johnston(1981)]{Hjellming1981}
Hjellming, R. M. \& Johnston, K. J. 1981, \apjl, 246, L141

\bibitem[Ho(2008)]{Ho2008}
Ho, L. C. 2008, \araa, 46, 475

\bibitem[Hulse \& Taylor(1975)]{Hulse1975}
Hulse, R. A. \&  Taylor, J. H. 1975, \apjl, 195, L51

\bibitem[Ji et al.(2021)]{Ji2021}
Ji, X., Lu, Y., Ge, J., Yan, C. \& Zhao, S. 2021, \apj, 910, 101


\bibitem[Kaspi et al.(2000)]{Kaspi2000}
Kaspi, S., Smith, P. S., Netzer, H., et al. 2000, \apj, 533, 631

\bibitem[Kato et al.(2008)]{Kato2008}
Kato, S., Fukue, J. \& Mineshige, S. 2008, Black Hole Accretion, Kyoto University Press.



\bibitem[King et al.(2013)]{King2013}
King, O. G., Hovatta, T., Max-Moerbeck, W. et al. 2013, \mnras, 436, L114

\bibitem[Kormendy \& Ho(2013)]{Kormendy2013}
Kormedny, J. \& Ho, L.\,C., 2013, \araa, 51, 511


\bibitem[Kovac{\v{e}}vi\'c et al.(2020a)]{Kovacevic2020a}
Kovac{\v{e}}vi\'c, A. B., Wang, J.-M., \& Popovi\'c, L. \v{C}. 2020, \aap, 635, A1

\bibitem[Kova{\v{c}}evi\'c et al.(2020b)]{Kovacevic2020b}
Kova{\v{c}}evi\'c, A. B., Songsheng, Y.-Y., Wang, J.-M. \& Popovi\'c, L. 2020, \aap, 644, 88 




\bibitem[Lacy et al.(2020)]{Lacy2020}
Lacy, M.; Baum, S. A.; Chandler, C. J. et al. 2020, \pasp, 132, 035001 

\bibitem[Leighly et al.(2016)]{Leighly2016}
Leighly, K. M., Terndrup, D. M., Gallagher, S. C., \& Lucy, A. B. 2016, \apj, 829, 4

\bibitem[Li et al.(2009)]{Li2009}
Li, Y.-R., Yuan, Y.-F., Wang, J.-M., et al. 2009, \apj, 699, 513
 
\bibitem[Li et al.(2016)]{Li2016}
Li, Y.-R., Wang, J.-M., Ho, L. C., et al. 2016, \apj,  822, 4

\bibitem[Li et al.(2019)]{Li2019}
Li, Y.-R., Wang, J.-M., Zhang, Z.-X., et al. 2019, \apjs, 241, 33 

\bibitem[Li et al.(2022)]{Li2022}
Li, Y.-R., Wang, J.-M., Songsheng, Y.-Y. et al. 2022, \apj, arXiv:2201.04470 
  
\bibitem[Liao et al.(2021)]{Liao2021}
Liao, W.-T., Chen, Y.-C., Liu, X. et al. 2021, \mnras, 500, 4025

\bibitem[Liu et al.(1999)]{Liu1999}
Liu, B. F., Meyer, F., Meyer-Hofmeister, E. \& Xie, G. Z. 1999, \apjl, 527, L17 

\bibitem[Liu \& Qiao(2020)]{Liu2022}
Liu, B. F. \& Qiao, E. 2022, iSceince,  25, 103544
  
\bibitem[Liu et al.(2019)]{Liu2019}
 Liu, T., Gezari, S., Ayers, M. et al. 2019, \apj, 884, 36
 
\bibitem[Luo et al.(2015)]{Luo2015}
Luo B., Brandt, W. N., Hall, P. B. et al., 2015, ApJ, 805, 122

 
\bibitem[Mahadevan(1997)]{Mahadevan1997}
Mahadevan, R. 1997, \apj, 477, 585
 
\bibitem[Manmoto(2000)]{Manmoto2000}
Manmoto, T. 2000, \apj, 534, 734

\bibitem[Martin et al.(2007)]{Martin2007}
Martin, R. G., Pringle, J. E. \& Tout, C. A. 2007, \mnras, 381, 1617

\bibitem[Mathews \& Brighenti(2003)]{Mathews2003}
Mathews, W. G. \& Brighenti, F. 2003, \araa, 41, 191

\bibitem[Meyer \& Meyer-Hofmeister(1994)]{Meyer1994}
Meyer, F. \& Meyer-Hofmeister, E. 1994, \aap, 228, 175

\bibitem[Meyer \& Meyer-Hofmeister(2002)]{Meyer2002}
Meyer, F. \& Meyer-Hofmeister, E. 2002, \aap, 392, L5



\bibitem[Milosavljevi\'c \& Merritt(2001)]{Milos2001}
Milosavljevi\'c, M. \& Merritt, D. 2001, \apj, 563, 34

\bibitem[Montuori et al.(2012)]{Montuori2012}
Montuori, C., Dotti, M., Haardt, F. et al. \mnras, 425, 1633

\bibitem[Muchotrzeb \& Paczy\'nski(1982)]{Muchotrzeb1981}
Muchotrzeb, B. \& Paczy\'nski, B. 1982, AcA, 32, 1

\bibitem[Narayan \& Yi(1994)]{Narayan1994}
 Narayan, R. \& Yi, I. 1994, \apjl, 428, L13

\bibitem[Narayan \& Yi(1995)]{Narayan1995}
 Narayan, R. \& Yi, I. 1995, \apjl, 452, 710 

\bibitem[Ni et al.(2022)]{Ni2022}
Ni, Q., Brandt, W. N., Luo, B. et al. 2022, \mnras, 511, 525

\bibitem[Noble et al.(2012)]{Noble2012}
Noble, S. C., Mundim, B. C., Nakano, H., et al. 2012, \apj, 755, 51

\bibitem[Noble et al.(2021)]{Noble2021}
Noble, S. C., Krolik, J. H., Campanelli, M., et al. 2021, \apj, 922, 175

\bibitem[Nyland et al.(2016)]{Nyland2016}
Nyland, K., Young, L. M., Wrobel, J. M. et al.2016, \mnras, 458, 2221

\bibitem[O'Neill et al.(2022)]{ONeill2022}
O'Neill, S., Kiehlmann, S., Readhead, A. C. S. et al. 2022, \apjl, 926, L35

\bibitem[Pacy\'nski \& Wiita(1980)]{Paczynski1980}
Pacy\'nski, B. \& Wiita, P. J. 1980, \aap, 88, 23

 

\bibitem[Paschalidis et al.(2021)]{Paschalidis2021} 
Paschalidis, V., Bright, J., Ruiz, M. et al. 2021, \apjl, 910, L26

\bibitem[Peters(1964)]{Peters1964}
 Peters, P. C. 1964, Phys. Rev. 136, B1224
 
\bibitem[Plotkin et al.(2015)]{Plotkin2015}
Plotkin R. M., Shemmer, O., Trakhtenbrot, B. et al., 2015, \apj, 805, 123 

\bibitem[Plotkinet al.(2010)]{Plotkin2010}
Plotkin, R. M., Anderson, S. F., Brandt, W. N., et al. 2010, \apj, 721, 562

\bibitem[Popovi\'c(2012)]{Popovic2012}
Popovi\'c, L. \v{C}, 2012, New A Rev., 56, 74

\bibitem[Popovi\'c, et al.(2021)]{Popovic2021}
Popovi\'{c}, L., Simi\'c, S.,  Kovac\v{e}vi\,c, A. \& Ili\'c, D. 2021, \mnras, 505, 5192

\bibitem[Rees et al.(1982)]{Rees1982}
Rees, M. J., Begelman, M. C. \& Blandford, R. D., 1982, \nat, 295, 17

\bibitem[Runnoe et al.(2017)]{Runnoe2017}
Runnoe, J. C., Eracleous, M., Pennell, A., et al. 2017, \mnras, 468, 1683

\bibitem[Sesana et al.(2012)]{Sesana2012}
Sesana, A., Roedig, C., Reynolds, M. T., \& Dotti, M. 2012, \mnras, 420, 860



\bibitem[Shakura \& Sunyaev(1973)]{Shakura1973}
  Shakura, N. I. \& Sunyaev, R. 1973, \aap, 24, 337 

\bibitem[Shapiro et al.(1976)]{Shapiro1976}
Shapiro, S. L., Lightman, A. P. \& Eardley, D. M. 1976, \apj, 204, 187


\bibitem[Shen \& Loeb(2010)]{Shen2010}
Shen, Y., \& Loeb, A. 2010, ApJ, 725, 249

\bibitem[Songsheng et al.(2019)]{Songsheng2019}
Songsheng, Y.-Y., Wang, J.-M., \& Li, Y.-R. 2019, \apj, 883, 184

\bibitem[Songsheng et al.(2020)]{Songsheng2020}
Songsheng, Y.-Y., Xiao, M., Wang, J.-M. \& Ho, L. C. 2020, \apjs, 247, 3

\bibitem[Stepney \& Guibert(1983)]{Stepney1983}
Stepney, S. \& Guibert, P. W. 1983, \mnras, 204, 1269


\bibitem[Schawinski et al.(2015)]{Schawinski2015}
Schawinski, K., Koss, M., Berney, S. \& Sartori, L. F., 2015, \mnras, 451, 2517



\bibitem[Valtonen et al.(2008)]{Valtonen2008}
 Valtonen, M. J., Lehto, H. J., Nilsson, K. et al. 2008, \nat, 452, 851

\bibitem[Vaughan et al.(2016)]{Vaughan2016}
 Vaughan, S., Uttley, P., Markowitz, A. G. et al. 2016, \mnras, 461, 3145
 
\bibitem[Volonteri et al.(2009)]{Volonteri2009}
Volonteri, M., Miller, J. M., \& Dotti, M. 2009, ApJ, 703, L86
 
\bibitem[Wang \& Li(2020)]{Wang2020a}
 Wang, J.-M., \& Li, Y.-R., 2020, RA\&A, 20, 160 
 
\bibitem[Wang \& Bon(2020)]{Wang2020b}
Wang, J.-M. \& Bon, E. 2020, \aap, 643, L9 

\bibitem[Wang et al.(2020)]{Wang2020c}
Wang, J.-M., Songsheng, Y.-Y., Li, Y.-R., Du, P. \& Zhang, Z.-X. 2020, NatAs, 4, 517

\bibitem[Wang et al.(2018)]{Wang2018}
Wang, J.-M., Songsheng, Y.-Y., Li, Y.-R., \& Yu, Z. 2018, \apj, 862, 171 

\bibitem[Wang et al.(2004)]{Wang2004}
Wang, J.-M., Watarai, K.-Y. \& Mineshige, S. 2004, \apjl, 607, L107

\bibitem[Wang \& Zhou(1999)]{Wang1999}
Wang, J.-M. \& Zhou, Y.-Y. 1999, \apj, 516, 420

\bibitem[Wang et al.(1999)]{Wang1999b}
Wang, J.-M., Szuszkiewicz, E., Lu, F. J. \& Zhou, Y.-Y. 1999, \apj, 522, 839

\bibitem[Welsh \& Horne(1991)]{Welsh1991}
Welsh, W. F. \& Horne, K. 1991, \apj, 379, 586

\bibitem[Wu et al.(2012)]{Wu2012}
Wu, J., Brandt, W. N., Anderson, S. F. et al. 2012, \apj, 747, 10

\bibitem[Yan et al.(2015)]{Yan2015}  
Yan, C.-S., Lu, Y., Dai, X., \& Yu, Q. 2015, \apj, 809, 117

\bibitem[Zhang et al.(2022)]{Zhang2022}
Zhang, P., Wang, Z., Gurwell, M. \& Wiita, P. J. 2022, \apj, 925, 207

\bibitem[Zilha\~o et al.(2015)]{Zilhao2015}
Zilha\~o, M., Noble, S. C., Campanelli, M., \& Zlochower, Y. 2015, PhRvD, 91, 024034

\end{thebibliography}
\end{document}